\documentclass[conference]{IEEEtran}
\usepackage{balance}

\usepackage{cite}
\usepackage{hyperref}

\usepackage{xcolor}
\usepackage{caption}
\ifCLASSINFOpdf
   \usepackage[pdftex]{graphicx}
\else
\fi
\usepackage{amsmath}
\usepackage{bm}

\ifCLASSOPTIONcompsoc
  \usepackage[caption=false,font=normalsize,labelfont=sf,textfont=sf]{subfig}
\else
  \usepackage[caption=false,font=footnotesize]{subfig}
\fi
\usepackage{csvsimple}

\begin{document}
%
\title{XbarSim: A Decomposition-Based Memristive Crossbar Simulator}

\author{\IEEEauthorblockN{Anzhelika Kolinko, Md Hasibul Amin, Ramtin Zand, Jason Bakos}
\IEEEauthorblockA{Department of Computer Science and Engineering, University of South Carolina, Columbia, SC 29208, USA\\
e-mail: akolinko@email.sc.edu, ma77@email.sc.edu, ramtin@cse.sc.edu, jbakos@cse.sc.edu
}
}

\maketitle

\begin{abstract}
Given the growing focus on memristive crossbar-based in-memory computing (IMC) architectures as a potential alternative to current energy-hungry machine learning hardware, the availability of a fast and accurate circuit-level simulation framework could greatly enhance research and development efforts in this field. This paper introduces XbarSim, a domain-specific circuit-level simulator designed to analyze the nodal equations of memristive crossbars. The first version of XbarSim, proposed herein, leverages the lower-upper (LU) decomposition approach to solve the nodal equations for the matrices associated with crossbars. The XbarSim is capable of simulating interconnect parasitics within crossbars and supports batch processing of the inputs. Through comprehensive experiments, we demonstrate that the XbarSim can achieve orders of magnitude speedup compared to HSPICE across various sizes of memristive crossbars. The XbarSim's full suite of features is accessible to researchers as an open-source tool.

\end{abstract}



%
\IEEEpeerreviewmaketitle

\section{Introduction}

Memristive crossbar arrays serve as vital components in analog processing-in-memory (PIM) and in-memory computing (IMC) architectures \cite{ankit2019puma,IMAC,shafiee2016isaac,AiMC, chi2016prime,PipeLayer,elbtity2023heterogeneous}. They offer significant acceleration for matrix-vector multiplication (MVM) operations in machine learning (ML) models by leveraging massive parallelism, analog computation, and minimizing data transfer overheads between memory and processor. Technologies utilized in crossbar-based IMC architectures include resistive random-access memory (RRAM) \cite{Yin2020HighThroughputIC}, phase-change memory (PCM) \cite{in-memory-PCM}, magnetoresistive random-access memory (MRAM) \cite{amin2022mram}, and conductive bridging random-access memory (CBRAM) \cite{Molas2019AdvancesIO}, among others.

As interest in IMC architectures grows, there remains a gap in fast and accurate circuit-level simulators capable of supporting hardware designers in implementing and validating their designs. Moreover, within the crossbar setting of IMC architectures, there is a wide array of design choices and hyperparameters that can be adjusted to fulfill particular design objectives. Hence, developing a simulation framework tailored to assist in the early-stage design deliberations for IMC circuits and architectures can prove highly advantageous.


The current crossbar simulation frameworks fall into two main categories: architecture-level \cite{mnsim,neurosim,rxnn} and circuit-level frameworks \cite{dpengine,imac-sim,cccs}. Architecture-level frameworks use analytical calculations to simulate the accuracy, power, and latency of crossbar outputs. Conversely, circuit-level simulators assess the entire circuit using circuit analysis techniques to derive accurate outputs for accuracy, power, and latency. While architecture-level simulations are faster, circuit-level simulations provide a more faithful representation of crossbar array behavior. Consequently, it is advisable for designers to conduct circuit-level evaluations followed by architecture-level assessments before making design decisions.

MNSim \cite{mnsim} is an architecture-level simulator employing analytical models to assess the performance of various elements within the crossbar. However, its validation reveals over 5\% error in power, energy, and latency calculations compared to full SPICE level simulations for a 3-layer fully connected neural network (NN).
NeuroSim \cite{neurosim} also offers analytical models for evaluating the power, area, and latency of crossbars.
RxNN \cite{rxnn} provides a faster model for crossbar simulation with non-idealities, but still uses some abstractions of non-ideal behaviour, lacking a full-circuit simulation.
As circuit-level simulations are essential for obtaining realistic results alongside architecture-level simulations, we propose a circuit-level simulation framework to enable researchers to conduct circuit-level evaluations for early-stage design decisions.

Some of the existing circuit-level simulation platforms are DPE \cite{dpengine}, CCCS \cite{cccs}, badcrossbar \cite{badcrossbar} and IMAC-Sim \cite{imac-sim}. In addition, \cite{chen2013} provided a model to estimate crossbar behavior, but lacks a simulation framework.
DPE \cite{dpengine}, a MATLAB-based simulator, focuses on finding an optimized mapping scheme for memristive crossbars considering non-ideal effects but lacks accuracy in interconnect resistance calculation. The CCCS \cite{cccs}, another MATLAB-based solver, suggests techniques for simplifying the evaluation of parasitic effects on the crossbars to speed up the simulation. However, it suffers from high error due to inaccurate estimation of voltage drops caused by parasitic resistance. Badcrossbar \cite{badcrossbar} is a python-based simulation framework, which uses python packages to solve the circuit equations. DPE, CCCS and badcrossbar use fixed resistances for evaluation, and while wire resistances are optimized to minimize errors, an accurate model for measuring parasitics remains absent. Recently, IMAC-Sim \cite{imac-sim}, a SPICE-based simulation framework, has been proposed which can automatically handle the SPICE simulations of crossbar-based IMC circuits. Although IMAC-Sim performs an accurate calculation of parasitics, the simulation time is high because of a full-circuit SPICE simulation. In this work, we propose a novel crossbar simulation framework, called XbarSim, that leverages the lower-upper (LU) decomposition technique to solve the nodal equations in a crossbar. XbarSim incorporates the parasitic resistances within crossbars and supports batch processing of the inputs.

\section{Methodology}


 

In this work, we consider the crossbar model shown in Fig. \ref{fig:crossbar}, comprised of horizontal word lines (WL) and vertical bit lines (BL), with a memristive device at the intersection between each WL and BL and a resistor between each pair of adjacent memristive devices to model the resistance of the wire. Each WL is driven with a voltage source to model the input vector, and the current through the termination of each BL corresponds to the output vector.

A mathematical description of the circuit is obtained by performing a nodal analysis, which produces a linear system of equations of the form $Gv = i$, where $G$ is a matrix of size $2 \times m \times n$ by $2 \times m \times n$  for a crossbar size of $m \times n$, $v$ is a set of $2 \times m \times n$ unknown internal node voltages, and $i$ are the currents flow through internal nodes into the ground node.

The circuit corresponding to an $m\times n$ crossbar contains $2mn$ nodes (i.e. two nodes per memristor), resulting in a matrix of size $2mn \times 2mn$ with $(4(n-2)+6)\times(4(m-2)+6)$ nonzero conductance entries. In the proposed approach, we use LU decomposition to pre-process the matrix to allow the node voltages to be calculated quickly.

\subsection{Circuit Model}


\begin{figure}[!t]
    \centering
    \includegraphics[width=3in]{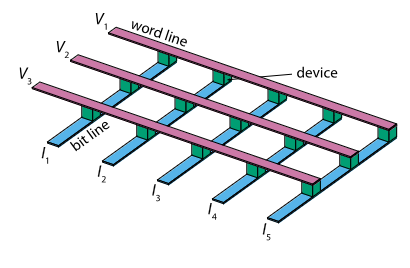}
    \label{fig:crossbar}
    \caption{The crossbar architecture \cite{badcrossbar}.}
    \label{fig:crossbar}
\end{figure}

\begin{figure*}[!t]
    \centering
    \includegraphics[width=6.5in]{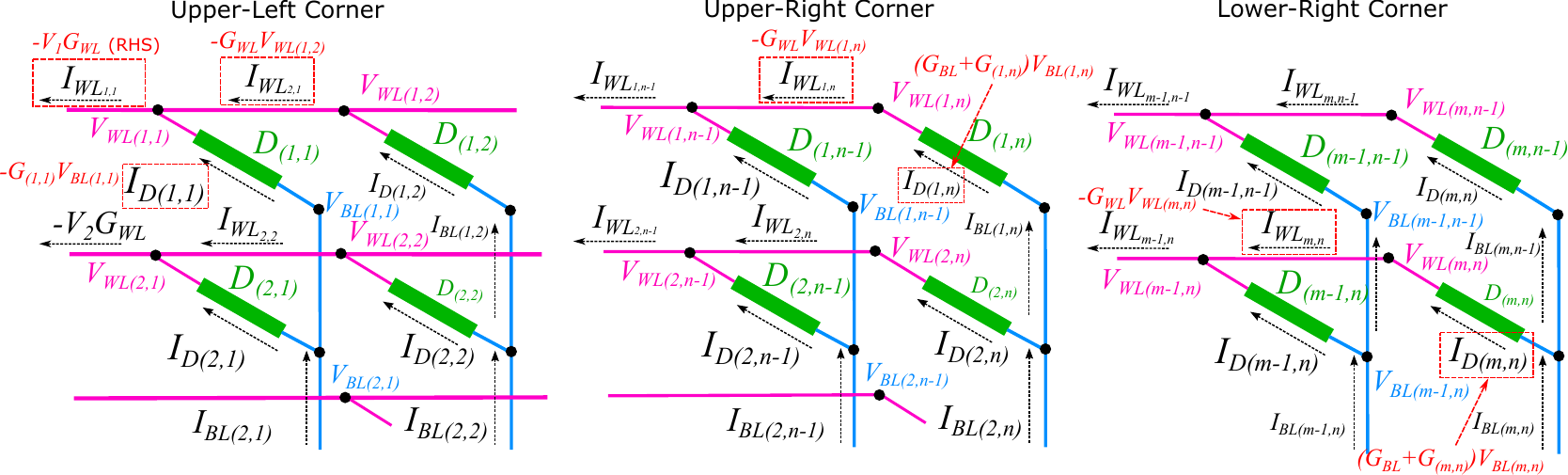}
    \caption{Notion used in crossbar matrix.}
    \label{fig:crossbar_notation}
\end{figure*}


Consider a crossbar of $m$ WLs and $n$ BLs. When including both memristors and a wire loss resistor between each pair of elements on the WLs and BLs, the circuit contains $3 \times m \times n$ resistors and $2 \times m \times n$ voltage nodes. The conductances of the resistors may be stored in a symmetric $2 \times m \times n$ by $2 \times m \times n$ matrix $\boldsymbol{G}$. The currents flowing in each node can be stored in a $2 \times m \times n$ vector $I$ (all but $n$ of which are zero). The vector of $2 \times n \times m$ node voltages, $V$ can be computed by solving the equation $\boldsymbol{G}V=I$. For constructing the matrix, assume the following notation:

\begin{enumerate}
    \item $V_i$: the voltage source at WL $i$,
    \item $D_{(i,j)}$: the memristor that is connected between WL $i$ and BL $j$,
    \item $G_{i,j}$: its conductance of $D_{(i,j)}$,
    \item $G_{BL}$: conductance of bit line (wire resistance),
    \item $G_{WL}$: conductance of word line (wire resistance),
    \item $V_{WL(i,j)}$: the voltage node of the WL-side of the memeristor,
    \item $V_{BL(i,j)}$: voltage node of the BL-side of the memristor,
    \item $I_{D(i,j)}$: current from $V_{BL(i,j)}$ to $V_{WL(i,j)}$,
    \item $I_{WL(i,j)}$: current from $V_{WL(i,j)}$ to $V_{WL(i,j-1)}$, and
    \item $I_{BL(i,j)}$: current from $BL_{i+1,j}$ to $V_{BL(i,j)}$.
\end{enumerate}

These notations are depicted in Fig. \ref{fig:crossbar_notation}, which shows three sections of the crossbar array, the upper left, upper right, and lower right, showing how the nodal equations are computed differently for the first row, first column, last row, and last column.


Kirchhoff's current law gives the following equations \cite{badcrossbar}:

\begin{equation}
    I_{WL(i,j)}-I_{WL(i,j+1)}-I_{D(i,j)}=0\ for\ j<n \label{eq:FKLForJsN}
\end{equation}

\begin{equation}
    I_{WL(i,j)}-I_{D(i,j)}=0\ for\ j=n \label{eq:FKLForJeqN}
\end{equation}

\begin{equation}
    I_{D(i,j)}-I_{BL(i,j)}=0\ for\ i=1 \label{eq:FKLForIeq1}
\end{equation}

\begin{equation}
    I_{D(i,j)}+I_{BL(i-1,j)}-I_{BL(i,j)}=0\ for\ i>1 \label{eq:FKLForIg1}
\end{equation}

When $j=1$ voltages at WLs are just input voltages:$V_{WL(i,j)}=V_{WL(i,1)}=V_i$ which is also shown in Eq.\ref{eq:current1}.

\begin{multline}
     (V_{WL(i,j)}-V_i)\times G_{WL}=(V_{BL(i,j)}-V_{WL(i,j)})\times G_{(i,j)}+\\+(V_{WL(i,j+1)}-V_{WL(i,j)})\times G_{WL}; \Rightarrow \\V_{WL(i,j)}\times (2\times G_{WL}+G_{(i,j)})+V_{WL(i,j+1)}\times (-G_{WL})+\\+V_{BL(i,j)}\times (-G_{(i,j)})=V_i\times G_{WL};\\for\ 1\leq i \leq m,\ j=1\label{eq:current1}
\end{multline}


Nodes corresponding to internal columns ($1<j<n$) and internal rows ($1<i<m$) can be computed as in Eqs. \ref{eq:current2} and \ref{eq:current3}. Other equations for node voltages can be derived similarly.

\begin{multline}
     (V_{WL(i,j)}-V_{WL(i,j-1)})\times G_{WL}=(V_{BL(i,j)}-V_{WL(i,j)})\times \\ \times G_{(i,j)}+(V_{WL(i,j+1)}-V_{WL(i,j)})\times G_{WL}; \Rightarrow \\V_{WL(i,j)}\times (2\times G_{WL}+G_{(i,j)})+V_{WL(i,j-1)}\times (-G_{WL})+\\+V_{WL(i,j+1)}\times (-G_{WL})+V_{BL(i,j)}\times s(-G_{D(i,j)})=0;\\for\ 1<i<m,\ 1<j<n\label{eq:current2}
\end{multline}

\begin{multline}
     (V_{BL(i,j)}-V_{WL(i,j)})\times G_{(i,j)}+(V_{BL(i,j)}- V_{BL(i-1,j)})\times G_{BL}=\\=(V_{BL(i+1,j)}-V_{BL(i,j)})\times G_{BL}; \Rightarrow\\
     V_{BL(i,j)}\times (2\times G_{BL}+G_{(i,j)})+V_{BL(i+1,j)}\times (-G_{BL})+\\+V_{BL(i-1,j)}\times (-G_{BL})+V_{WL(i,j)}\times (-G_{(i,j)});\\ for\ 1<i<m,\ 1<j<m\label{eq:current3}
\end{multline}

\subsection{Matrix Formulation}

The resulting equations form a sparse system of linear equations. This system is equivalent for row and column permutations, since row exchanges correspond to a change in the order of equations, while column exchanges correspond to a change in the order of summands in each equation, assuming that corresponding permutations are performed on the vector of unknowns and the right-hand side vector of known values.

\begin{figure}[!t]
\centering
\includegraphics[width=3.45in]{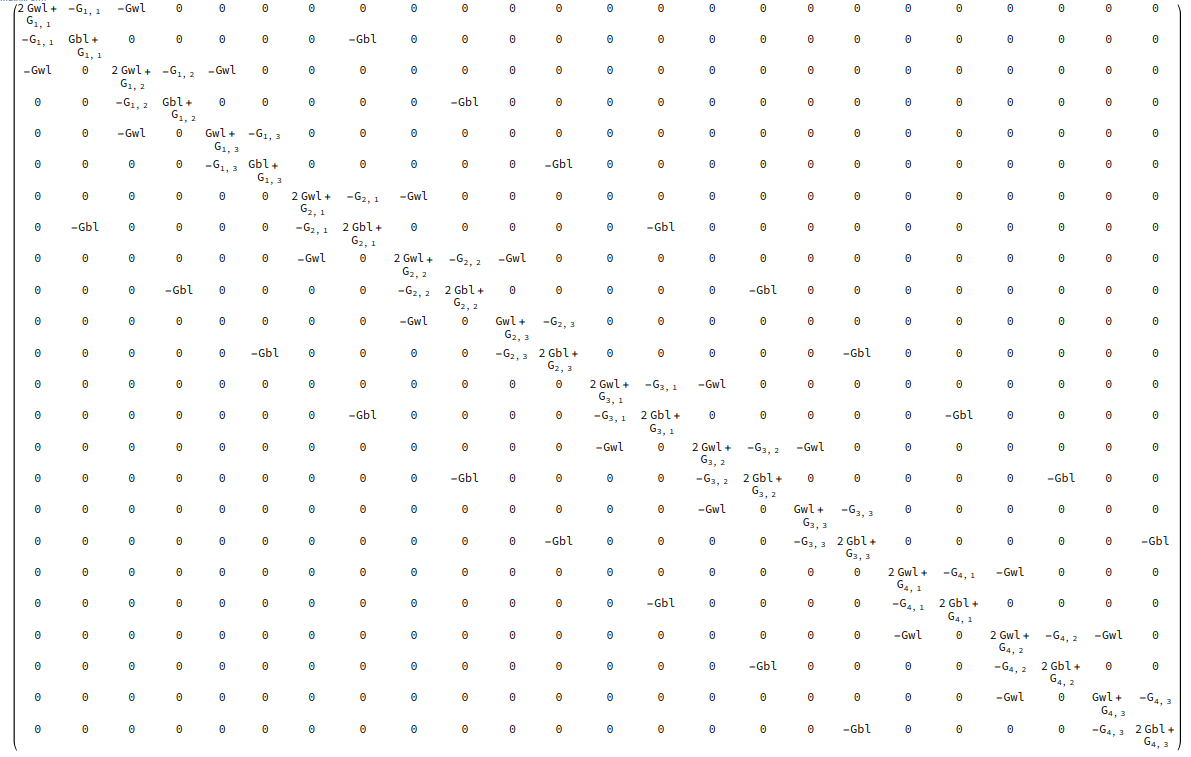}
\caption{Example of a matrix formed from nodal equations for a 4 by 3 crossbar.}
\label{fig:bd_mx}
\end{figure}

One can use a permutation of columns $\{V_{WL(1,1)}, V_{BL(1,1)}, V_{WL(1,2)}, V_{BL(1,2)}, ...,V_{WL(m,n-1)},\\V_{BL(m,n-1)}, V_{WL(m,n)}, V_{BL(m,n)}\}$ that corresponds to the vector of unknown nodal voltages. To make the resulting matrix banded diagonal, one can use such a permutation of rows that the matrix value $M_{i,j}$ at the diagonal would be the corresponding unknown coefficient on the right-hand side of the system of linear equations. Hence, for selected column permutation, corresponding row permutation would be nodal equations corresponding to equations $\{$\ref{eq:FKLForJsN} ($i=1,j=1$),  \ref{eq:FKLForIeq1} ($i=1,j=1$), \ref{eq:FKLForJsN} ($i=1,j=2$), \ref{eq:FKLForIg1} ($i=1,j=2$),..., \ref{eq:FKLForJsN} ($i=m,j=n-1$), \ref{eq:FKLForIg1} ($i=m,j=n-1$), \ref{eq:FKLForJeqN} ($i=m,j=n$), \ref{eq:FKLForIg1}$\}$. An example of a matrix formed this way for $m=4,n=3$, can be seen in Fig. \ref{fig:bd_mx}.



\subsection{Solution Cost}

There are numerous potential methods for solving the resultant system, including iterative methods and decomposition methods. Given that the matrix is large and sparse, iterative methods are an obvious approach. However, 
iterative methods are sensitive to the initial guess, which 
is not practical for this application. On the other hand, decomposition methods 
can also diverge in cases where several rows (or, equivalently, columns) are close to linearly dependent. However, since we are modeling a real physical system, the resulting matrix should have a solution and, hence, it is expected to have linearly independent rows (or, equivalently, linearly independent columns).

For this reason, we use an LU decomposition-based solver, in which $\boldsymbol{G}V = I$ is transformed into $\boldsymbol{LU}V = I$, where $\boldsymbol{L}$ is a lower triangular matrix with ones on its diagonal and $\boldsymbol{U}$ is an upper triangular matrix.

Substituting $y=\boldsymbol{U}V$ into the equation gives $\boldsymbol{L}y=I$. The solution for the vector $y$ can be calculated with no more than $L_{nnz}$ multiply-accumulates, where $nnz$ is the number of non-zero entries. The solution for the vector $x$ can be calculated by substituting $y$ into $\boldsymbol{U}V=y$ and solving for $V$, which requires no more than $U_{nnz}$ scalar multplies, adds, and divides.

There are several algorithms for performing LU decomposition. Matlab's default built-in LU decomposition uses a variant of Gaussian elimination \cite{LU_Matlab}. When using Gaussian elimination for LU decomposition, the computational cost is $O(k^3)$ multiple-accumulates for a $k\times k$ matrix. For a the banded diagonal $k\times k$ matrix with bandwidth $m$, the cost is $O((\frac{m-1}{2})^2\times k)$.

The workload required to perform an LU decomposition on sparse matrix depends on the sparsity of the matrix but is bounded by the $A_{nnz}^2$ multiply-accumulates, although in practice the execution time may be dominated by the overheads required by reading and writing sparse matrix storage format.

\section{Results}
\begin{figure*}[!t]
    \centering
    \subfloat[Nonzero elements of matrix generated for a $50\times2$ crossbar]{\includegraphics[width=2.8in]{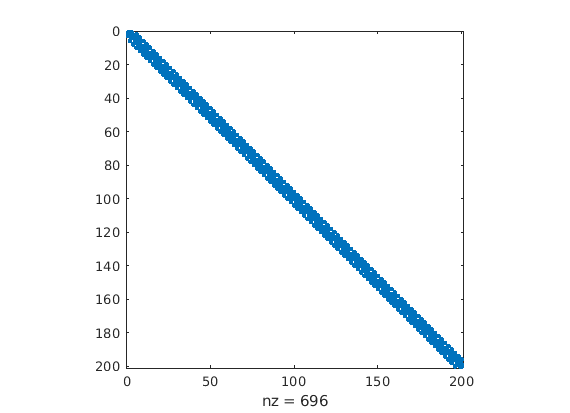}%
    \label{fig:5by2}}
    \hfil
    \subfloat[Nonzero elements of matrix generated for a $2\times50$ crossbar]{\includegraphics[width=2.8in]{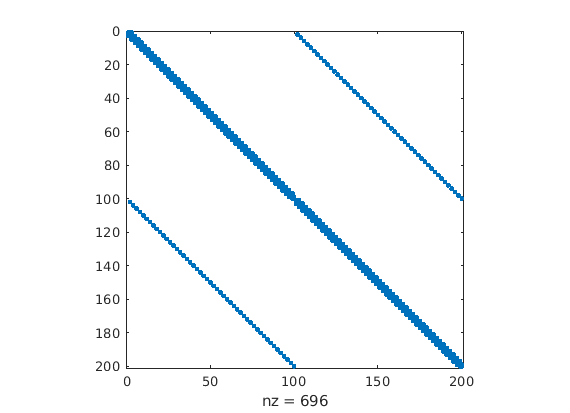}%
    \label{fig:2by50}}
    \caption{Positions of nonzero elements in generated matrix.}
    \label{fig:mbyn_spy}
\end{figure*}

For the simulations, we implemented our proposed LU decomposition-based method in MATLAB and compared the results with HSPICE 2019.06-2 serving as the baseline. 
Here, we study two scenarios: when the simulation is performed for a crossbar with one set of input voltages, and when multiple sets of input voltages are applied in batch mode.

\subsection{System architecture}

Our software generates a SPICE circuit corresponding to a crossbar circuit of a given size. In the generated file, the memristor conductances and the input voltages are randomly generated, and the parasitic resistors are set at 5 Ohms each \cite{parasitics,amin2022xbar}. We perform a DC steady-state analysis of the circuit using HSpice 2019.06-2 with the option  ``.option post=2 POST\_VERSION=2001'' to store the output nodal voltages in an ASCII output file. Batch runs are carried out by associating a ``.DATA'' values with the input voltage sources. HSPICE performs a DC analysis for each of a series of input voltage values.


For data-driven analysis with batch size sets of input voltages, which should translate into the right-hand side vector of the matrix equation, one needs to decompose the matrix only once for the batch. Each simulation was run 10 times to gather the mean solve time. Batch sizes of 1, 2, 10, 100, and 1000 are investigated.



\subsection{Simulation Time}\label{ss:batch_1}

To compare the proposed solver with HSpice, a set of $m \times n$ crossbars is generated, in which $m, n \in \{2,4,8,16,32,64,128,256\}$.
Each simulation is performed 10 times, and the mean execution time is computed. 

\begin{figure}[!t]
\centering
\includegraphics[width=3.4in]{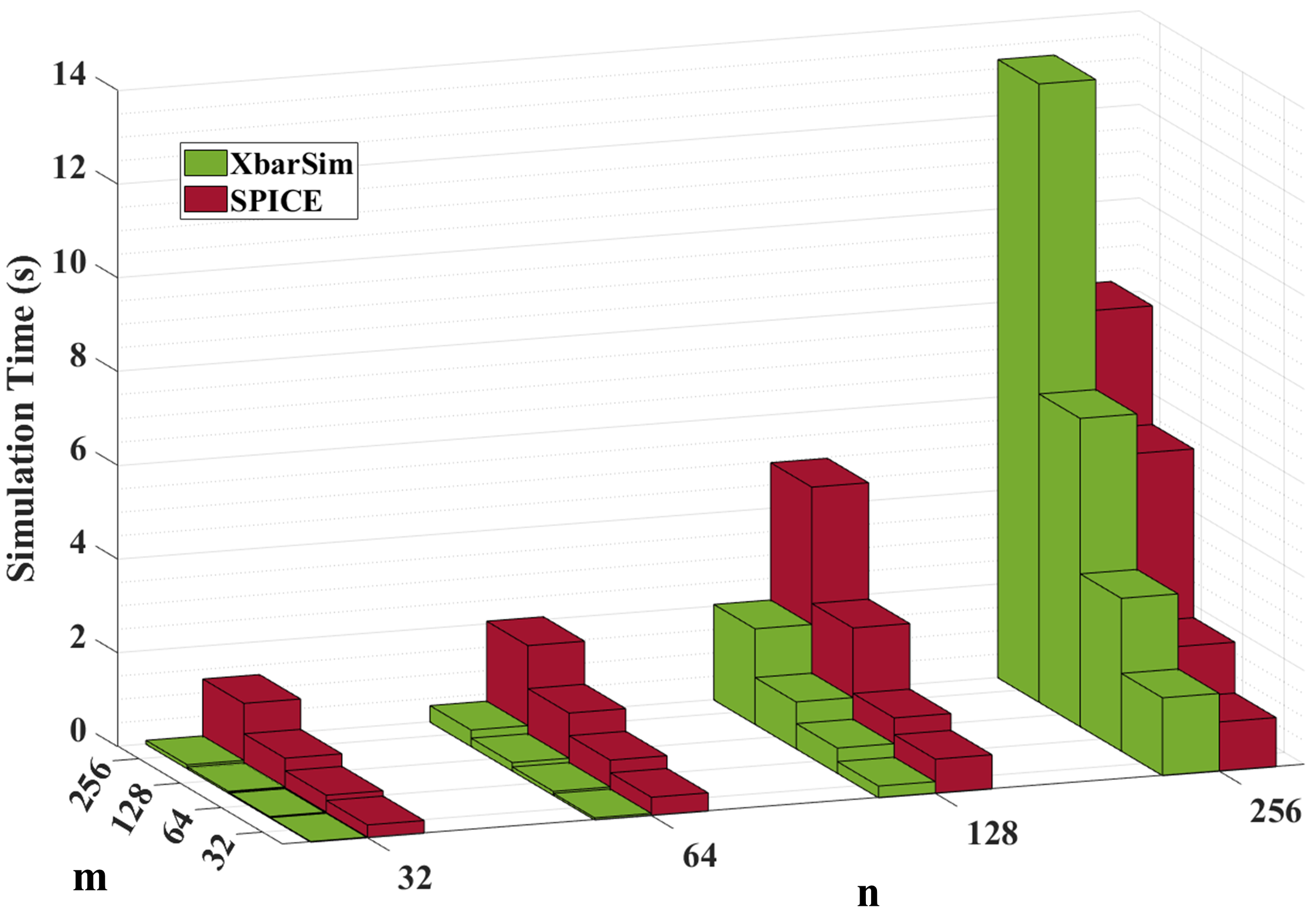}
\caption{Simulation time for various $m\times n$ crossbars}
\label{fig:crossbar_time}
\end{figure}

Figure \ref{fig:crossbar_time} shows the simulation time for our simulator and SPICE with various crossbar sizes. The results show an increase in simulation time for larger crossbar sizes. For example, the simulation time for a $32 \times 32$ crossbar is 5 ms, which increases to 85 ms for a $32 \times 256$ crossbar and 1.66 s for a $256 \times 32$ crossbar. The simulation time in our simulator is lower than SPICE except when the number of columns (n) is set to 256. For example, while the simulation time of 270 ms for a $32 \times 32$ crossbar in SPICE is higher than the time in our simulator, for a $32 \times 256$ crossbar, the SPICE simulation time is 1.05 s, which is lower than the 1.66 s for our simulator.\\

A possible explanation for this observation is that although for $m\times n$ and for $n \times m$ crossbars number of nonzero elements are the same, these nonzero elements are distributed differently. As one can see from Fig. \ref{fig:mbyn_spy} if $m$ is significantly larger than $n$ that does not significantly influence the overall bandwidth of the resulting matrix compared to that of its transposed form. In contrast, if $n$ is significantly larger than $n$, one of the lateral diagonals of the matrix is located further away from the main diagonal, thus increasing its bandwidth and correspondingly potentially increasing the decomposition time.

\subsection{Partitioning}
\begin{figure}[!t]
\centering
\includegraphics[width=3.4in]{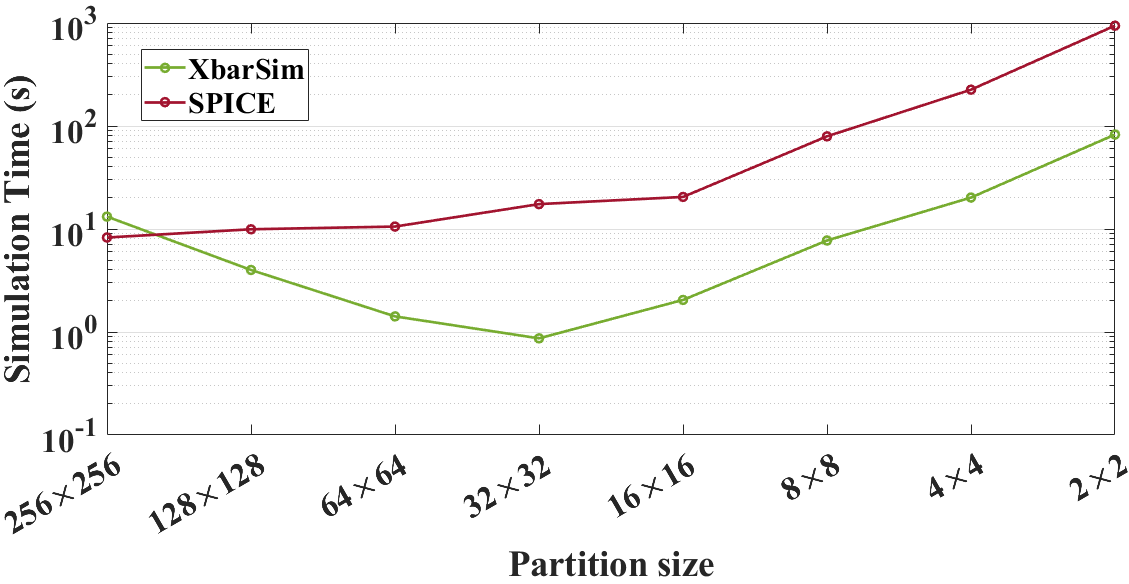}
\caption{Simulation time for $256\times256$ crossbar partitioned with different crossbar sizes.}
\label{fig:partition}
\end{figure}

As larger crossbars need a longer simulation time, we partition the larger crossbars into various smaller crossbar sizes. Figure \ref{fig:partition} shows the simulation time for a $256 \times 256$ crossbar partitioned into various smaller crossbar sizes. The results show that there is an optimum partition size where the simulation time is minimized. As shown in Fig. \ref{fig:partition}, for a $256 \times 256$ crossbar, we obtain the minimum simulation time 870 ms, when we partition it into $32\times32$ crossbars, while the simulation time is 13.19s for an equivalent unpartitioned simulation.

\subsection{Batch Processing}
\begin{figure}[!t]
\centering
\includegraphics[width=3.4in]{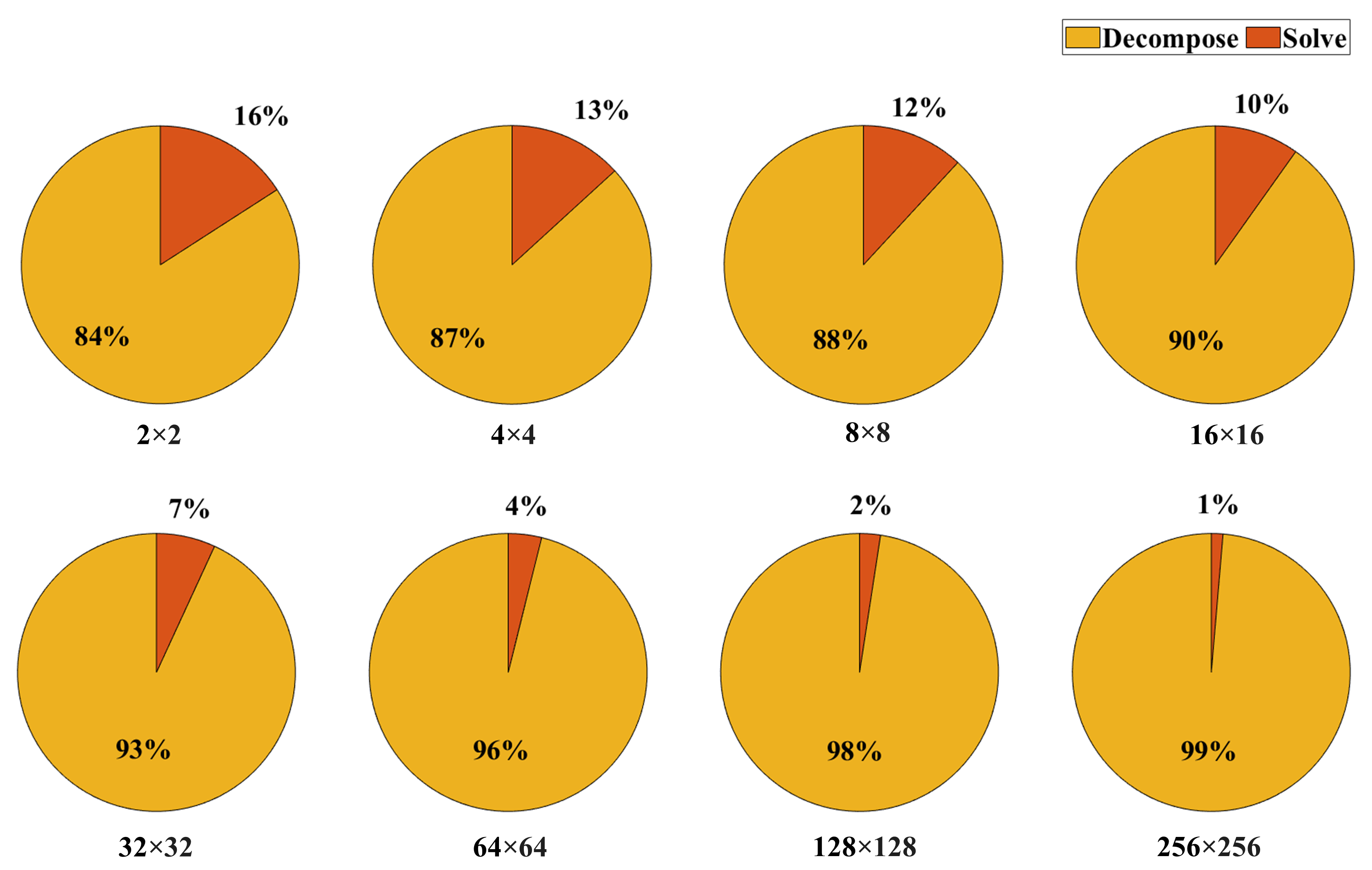}
\caption{Ratio of decomposition and solve time for various crossbar sizes.}
\label{fig:pie_symmetric}
\end{figure}

\begin{figure}[!t]
\centering
\includegraphics[width=3.4in]{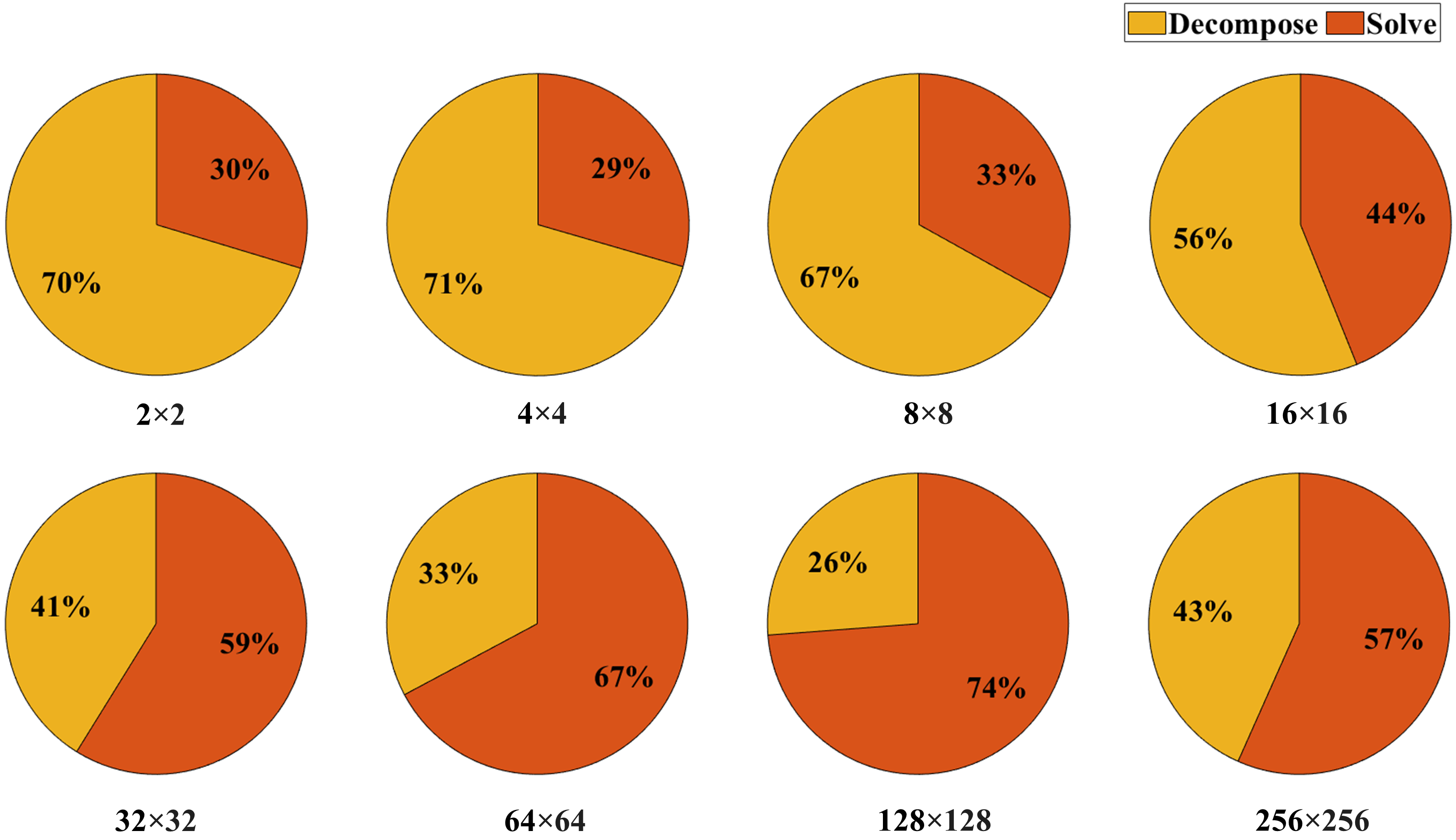}
\caption{Ratio of decomposition and solve time for various crossbar sizes with a batch size of 100.}
\label{fig:pie_batch}
\end{figure}

In Fig. \ref{fig:pie_symmetric}, we show the decomposition and solve ratio of our simulator for various crossbar sizes. From Fig. \ref{fig:pie_symmetric}, we observe that most of the simulation time is spent in the decomposition phase. Hence, we batch multiple input samples in a single simulation, where we decompose the matrix once and run the solve phase for multiple input samples. We tested our solver with batch sizes of 2, 10, 100, and 1000. As an example, Fig. \ref{fig:pie_batch} shows the decomposition to the solve time ratio for various crossbar sizes with a batch size of 100. The figure illustrates how processing inputs in batches leads to solve time becoming the predominant factor in simulation time.

\begin{figure}[!t]
\centering
\includegraphics[width=3.4in]{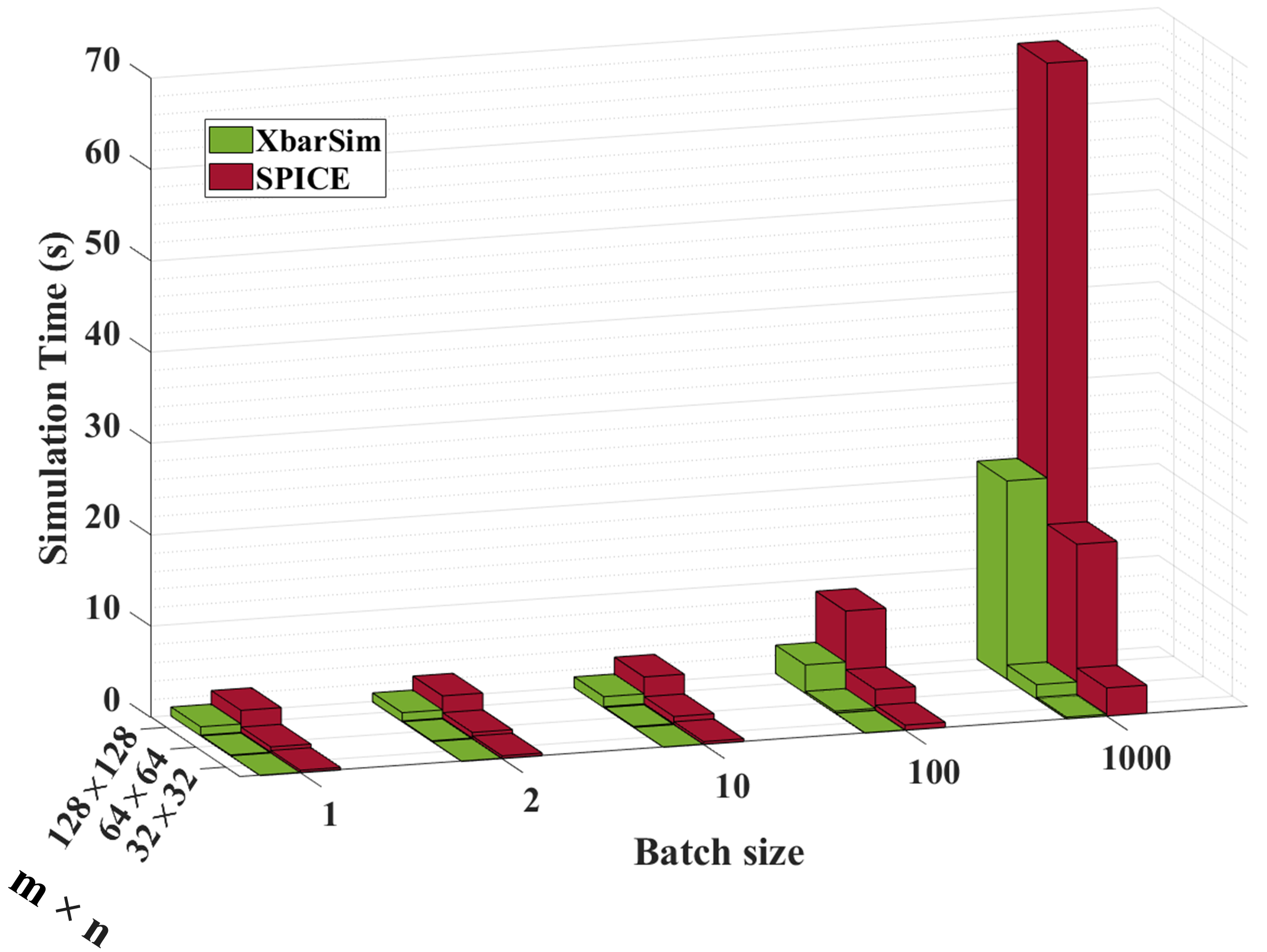}
\caption{Simulation time for various crossbars with different batch sizes.}
\label{fig:batch_time}
\end{figure}

\begin{figure}[]
\centering
\includegraphics[width=3.4in]{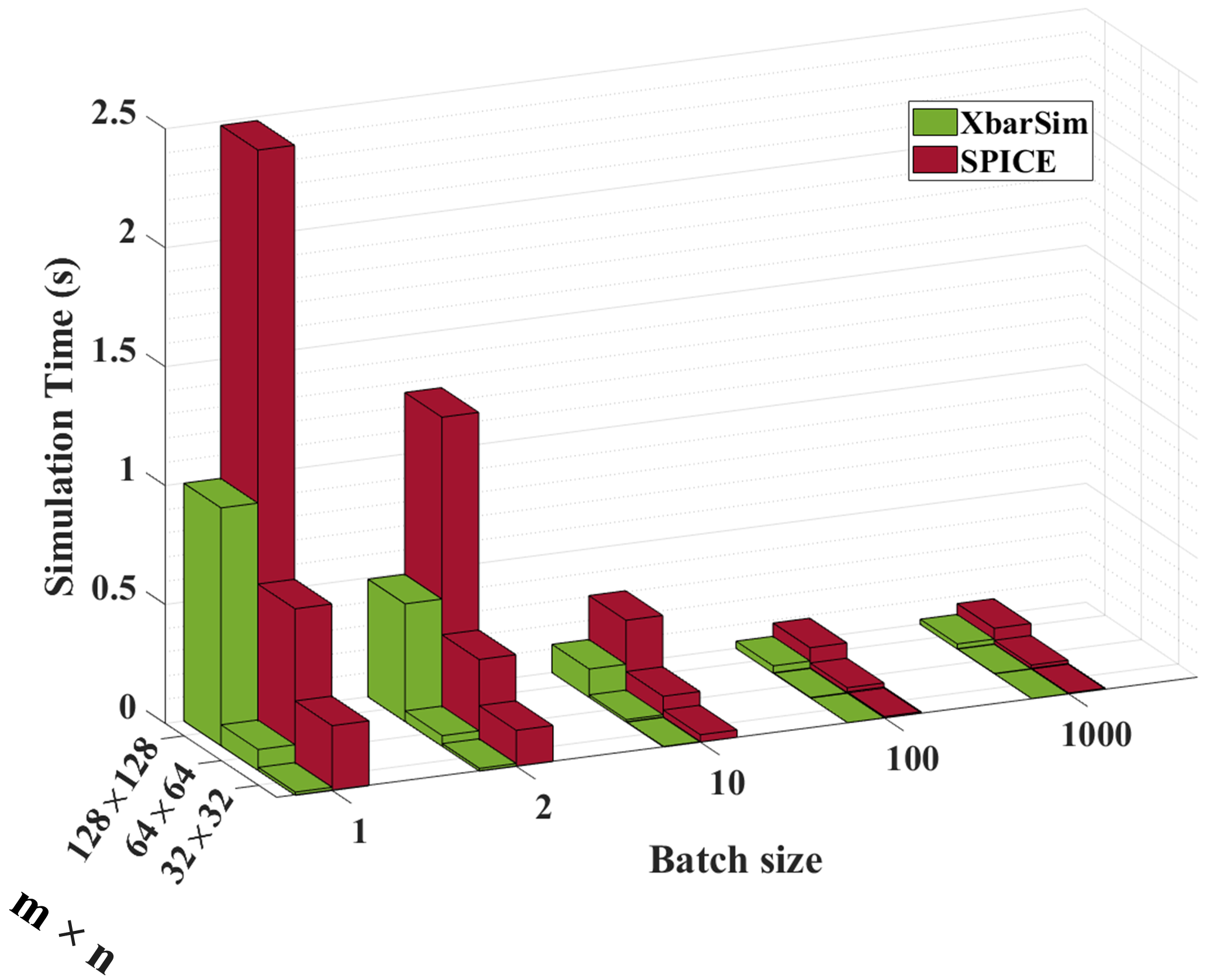}
\caption{Simulation time per input sample for various crossbars with different batch sizes.}
\label{fig:batch_sample}
\end{figure}

Fig. \ref{fig:batch_time} shows the results of batching multiple input data. For a $128\times128$ crossbar, the simulation time is 1 s without batching. With a batch size of 1000, the simulation time is 21.71 s, a significant improvement considering the simulation time per input sample. Fig. \ref{fig:batch_sample} shows the simulation time per input sample for various crossbar sizes. The simulation time per sample improves with increasing batch size. For example, with the $128\times128$ crossbar, the simulation time per sample is 1 s for a batch size of 1. The time reduces to 500 ms for a batch size of 2 and further decreases to 22 ms for a batch size of 1000.

\section{Conclusion}

In this paper, we introduced XbarSim, a specialized simulation framework tailored for solving matrices associated with memristive crossbar nodal equations. To solve the matrices, we employ decomposition-based methods to mitigate the sensitivity to initial guesses in iterative methods. Extensive simulations and comparisons with the HSPICE circuit simulator are conducted across crossbars of varying dimensions. Our findings reveal significant performance improvements for XbarSim compared to HSPICE, particularly for crossbars with fewer than 256 columns. To address performance slowdowns with 256-column crossbars, we equipped XbarSim with partitioning capabilities, achieving a 9.5$\times$ speedup compared to HSPICE by partitioning a 256$\times$256 crossbar into sixteen 32$\times$32 segments. Moreover, we explore the impact of batch processing on simulation time, demonstrating up to a 45$\times$ speedup per sample simulation for a 128$\times$128 crossbar with a batch size of 1,000. The speedups achieved by XbarSim provide several opportunities for future work, including extending XbarSim to support complete neural network simulations and integrating it with multi-objective optimization algorithms for optimizing ML workload mapping on crossbar-based IMC architectures.






%



\section*{Acknowledgment}
This work is supported in part by the National Science Foundation (NSF) under grant number 2409697.

\balance
\bibliographystyle{IEEEtran}
\bibliography{refs}

\end{document}